# Music Generation using Human-In-The-Loop Reinforcement Learning

Aju Ani Justus
*School of Computer Science*
*University of Birmingham*
Birmingham, United Kingdom
axa1943@alumni.bham.ac.uk

*Abstract*—This paper presents an approach that combines Human-In-The-Loop Reinforcement Learning (HITL RL) with principles derived from music theory to facilitate real-time generation of musical compositions. HITL RL, previously employed in diverse applications such as modelling humanoid robot mechanics and enhancing language models, harnesses human feedback to refine the training process. In this study, we develop a HILT RL framework that can leverage the constraints and principles in music theory. In particular, we propose an episodic tabular Q-learning algorithm with an epsilon-greedy exploration policy. The system generates musical tracks (compositions), continuously enhancing its quality through iterative human-in-the-loop feedback. The reward function for this process is the subjective musical taste of the user.

## I. INTRODUCTION

The aim of the algorithm presented in this paper is to generate musical compositions without dependence on pre-existing musical data, thus contributing to the domain of computer-aided music generation. This paper seeks to combine principles rooted in music theory with reinforcement learning techniques, contributing to the studies on algorithmic music generation free from data dependencies.

The motivation behind this paper stems from the desire to equip musicians and composers with a valuable tool for crafting original music compositions, free from external influences by either open-source or copyrighted musical sources. Additionally, it addresses a recognised knowledge gap in the exploration of Human-In-The-Loop Reinforcement Learning (HITL RL) for music generation. Due to the subjective nature of music generation, enabling users to tailor the model to their unique preferences and personal taste would be beneficial. Furthermore, the prospect of sharing these user-tailored models holds the potential for musicians to openly exchange their models for inspiration, thereby fostering collaborative innovation within the field.

The significance of this paper lies in the possible advancement of the field of computer-aided music generation that avoids dependency on existing data. The unexplored territory of non-data-driven, algorithmic, and user-guided music generation necessitates further inquiry, with this paper potentially serving as the cornerstone for subsequent studies within the realm of non-data-driven music generation frameworks that combine HITL and RL. Moreover, although HITL and RL have been independently applied in the context of music generation, a pronounced knowledge gap emerges upon their integration (HITL RL).

### A. Summary

An overview of the proposed framework is depicted in Fig. 1. This approach consists of two main components, with user interaction at its core. The first component, known as the MusicGenerator, is responsible for creating music tracks. Subsequently, the user rates these tracks on a scale from 1 to 10. The second component, the HITL_RL_Agent, uses an episodic tabular Q-learning approach based on an epsilon-greedy exploration policy to adjust melodies based on the user rating. The Graphical User Interface (GUI) serves as a tool for user input, enabling music generation configuration, model training, and model management.

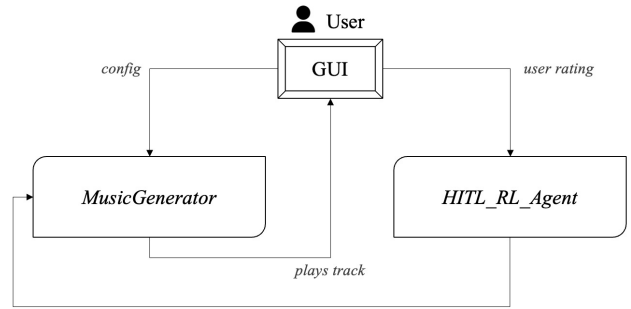

Fig. 1. Overview of the main components and user interaction.

## II. BACKGROUND

This paper explores the intersection of reinforcement learning (RL) and Human-In-The-Loop (HITL) algorithms in the context of music generation. In this section, we provide a comprehensive overview of existing literature and highlight the pivotal role of HITL RL in addressing knowledge gaps within the field.

### A. Reinforcement Learning

Music generation with RL has gained attention for its ability to produce original compositions. Hong *et al.* [1] introduced a deep reinforcement learning method for single-track MIDI music generation, emphasising the incorporation of existing MIDI music data for training and enhancing harmony. Similarly, Jin *et al.* [2] proposed a framework for metaverse concert generation, utilising RL techniques and Transformer-XL [3] music generation network trained on the POP909 dataset [4], underlining the significance of existing music data.

Reese *et al.* [5] explored tonal music generation through geometric topologies and RL models. Their approach constructed geometric networks of chords based on existing chord data, utilising RL models for learning chord progressions, highlighting the importance of available chord progressions and data related to musical scales. Chen *et al.* [6] presented an automatic composition method for Guzheng music, relying on existing Guzheng music pieces for training with RL techniques to capture the unique characteristics of Guzheng music.

*B. Human-In-The-Loop Algorithms*

The term HITL algorithms refers to computational systems that integrate human input and feedback as essential components in their learning or decision-making processes. In the context of music generation, HITL algorithms aim to combine the computational power of algorithms with the creative insights and preferences of human users.

The adoption of HITL algorithms in music generation has been thoroughly explored, e.g., refer to [7, 8, 9]. Bryden [7] introduced a HITL evolutionary algorithm for data-driven music, emphasising user-guided evolution. Pei [8] conducted a survey on sound and music composition using interactive evolutionary computation, showcasing the role of HITL in auditory design. Tavakoli's HARMONY project [9] demonstrated the power of human-centred data collection in complex tasks. This collective evidence underscores HITL as a versatile methodology, promising more nuanced and robust algorithmic solutions across diverse applications.

*C. Human-In-The-Loop Reinforcement Learning*

Contrary to traditional RL methodologies, HITL RL engages users throughout the learning trajectory, harnessing human expertise to refine and elevate algorithmic outputs. This nuanced approach is particularly relevant to creative domains, where the subtleties of artistic expression necessitate continuous human guidance.

The literature on HITL RL across varied applications attests to its adaptability and efficacy. Alamdari *et al.*'s pioneering work [10] in personalised hearing aid compression exemplifies the algorithm's accommodation of individual user preferences. Luo *et al.*'s exploration [11] into continuous action spaces underscores its potential to augment learning in intricate environments. Studies by Reese *et al.* [5] and Wu *et al.* [12] underscore the symbiosis of human intelligence and machine learning, showcasing the versatility of HITL RL across domains. Human-agent teaming amplifies adaptability, efficiency, and performance, especially in creative spheres where subjective discernment assumes paramount importance. Nonetheless, the application of HITL RL in music generation remains uncharted territory, which is a surprising knowledge gap.

*D. Knowledge Gaps*

The fields of HITL algorithms and RL have individually made significant contributions to creative applications. However, a noteworthy knowledge gap exists in their combined application (HITL RL), specifically in the realm of music generation. While HITL algorithms and RL have been employed separately for music generation, "Music Generation using Human-In-The-Loop Reinforcement Learning" remains relatively unexplored. Furthermore, using HITL RL has the potential to overcome the knowledge gap about the dependency of RL algorithms on existing music data.

Another emerging gap in the field of RL-based music generation pertains to the dependency on existing music data. While the reviewed studies showcase the power of reinforcement learning in generating music, they consistently underscore the importance of pre-existing music data for training and optimisation. This reliance on existing data raises intriguing questions about the extent to which RL can create music autonomously without data dependencies. Further exploration of this knowledge gap could pave the way for innovative approaches in music generation that reduce reliance on pre-existing musical compositions.

In conclusion, the adoption of HITL RL in music generation represents an innovative step toward personalised and emotionally resonant musical experiences. By bridging the gap between technology and artistic expression, HITL RL has the potential to expand on the research on music generation without music data dependency. This uncharted territory holds promise for shaping the future of music generation.

III. FORMULATION OF THE SOLUTION

*A. Solution Formulation and Algorithm*

**Q-Learning:** Q-Learning is a temporal-difference (TD) algorithm where a Markov Decision Process (MDP), a framework for sequential decision problems [13], is considered. An MDP is defined as *MDP (S, A, P(s', s, a), R(s', s, a))*, where *S* is the set of possible states, *A* is the set of possible actions, $P(s', s, a) = P(s_{t+1}=s' | s_t=s, a_t=a)$ is the transition model that maps a new state *s'* from state *s* through action *a* via one-step state transition dynamics, and *R* describes the reward.

The solution is built upon the concept of a *track array*, a 2-dimensional array representing the musical composition, with melody and percussion elements. The *track array* is built on established music theory principles, initialised based on user-specified parameters—base note, track length, and type of musical scale from {major, minor, diminished}. A scale is then generated based on these parameters (e.g., $C_4$ as base note and major scale-type selection yielding a $C_4$ major scale). Melodic note pitches are randomly selected from this scale. Note duration and percussion elements are also randomly generated for melody and percussion arrays.

*track array = [melody array, percussion array]*
*melody array = [(pitch_1, duration_1), (pitch_2, duration_2), ...]*
*percussion array = [percussion_pitch_1, percussion_pitch_2, ...]*

Within this framework, the Q-Learning algorithm refines the *track array*, adapted for discrete action spaces *S*. Bellman's Equation guides decision-making, with the Q-value *Q(s, a)* representing the cumulative reward for a specific action *a* in a given state *s*. The learning rate *α* controls the step size of Q-value adjustments, and immediate reward *R(s, a)* signifies the user rating on a scale from 1 to 10. The discount factor *γ* balances consideration of immediate and future rewards. The state resulting from an action *a* in state *s* is denoted as *s'*, and *a'* encompasses possible actions in the resulting state. The action space *A* includes discrete alterations to the *track array*: $A=\{0,1,2,3,4,5\}$, where 0 increases the pitch by 1, 1 decreases the pitch by 1, 2 increases note duration by 0.25 (capped at 1), 3 decreases note duration by 0.25 (capped at 0.25), 4 changes the pitch of a percussion instrument, and 5 removes a note. The Q-value update rule is defined as:

$$Q(s, a) \leftarrow Q(s, a) + \alpha [R(s, a) + \gamma * max(Q(s', a')) - Q(s, a)]. \quad (1)$$

*B. Exploration Strategies*

The Epsilon-Greedy strategy balances exploration and exploitation in the agent's decision-making.

- With probability ε, the agent selects a random action (exploration).

- With probability 1-ε, the agent selects the action with the highest estimated value based on its current Q-values (exploitation).

The exploration-exploitation trade-off is an open problem in RL. Within the context of music generation, exploration translates into encouraging random actions for the discovery of new compositions, whereas exploitation leverages learned knowledge for refined compositions.

*C. Training Process*

The training process consists of a minimum of 10 episodes, exposing the agent to diverse musical contexts. Each episode begins by generating a *track array* solely based on user inputs and principles of music theory, as detailed in Section III-A, with no external data involved. During each step, the *track array*, representing the discrete state *s*, undergoes meticulous modifications guided by the RL agent's actions, employing the Epsilon Greedy strategy with an initial exploration-exploitation parameter ε set to 0.5. At each step of every episode, the music is generated based on the *track array* and played back to the user, who rates it on a scale of 1-10. User ratings serve as immediate rewards, influencing the agent's decisions. The Bellman's Equation, incorporating the specified values of $α=0.1$ and $γ=0.9$, is applied in each episode to update Q-values. This iterative process refines the model's understanding of aesthetically pleasing compositions in a user-centric manner.

Users have the option to extend training beyond the initial 10 episodes, allowing the model to evolve. As users invest time in a large number of episodes, the model essentially refines its Q-Table, becoming better suited to generate compositions tailored to individual tastes. The ability to save and load models underscores the user-driven nature of the training process, emphasising ongoing adaptation to evolving user preferences while upholding the model's integrity and authenticity in music creation. Crucially, no external data or existing music is provided to the agents. The model relies on randomisation within the scale generated based on user inputs for *track array* initialisation, ensuring creative autonomy and preventing biases from external influences.

*D. Evaluation Metrics*

In this paper, we employ specific evaluation metrics to assess the quality of the generated music, with testing users providing ratings on a scale of 1-5 for each of the metrics—Musicality, Novelty, and Coherence. These metrics are self-defined for the purpose of evaluating the algorithmic generated music, drawing inspiration from established principles in music theory while tailored to the objectives of this study.

Musicality: Assessing the degree to which the generated compositions sound musically pleasing and coherent. It gauges the alignment with musical theory principles, emphasising melody, harmony, dynamics, and tonal balance. Beyond technical correctness, high musicality signifies compositions that evoke emotion and demonstrate refined artistic qualities, ensuring a pleasing and enjoyable listening experience.

Novelty: Measuring the uniqueness and originality of the generated music compared to pre-existing compositions. This metric identifies unique patterns, melodic structures, and innovative combinations of musical elements, showcasing a departure from conventional musical norms. High novelty indicates a successful push of creative boundaries, offering a fresh and distinct listening experience with compositions that stand out for their unique contributions to the musical landscape.

Coherence: Evaluating the structural coherence and consistency within generated compositions. It focuses on the seamless integration of musical elements like harmony, rhythm, and transitions, ensuring a unified and well-organised musical piece. A coherent composition exhibits logical progressions, with each section flowing naturally into the next. High coherence signifies structurally sound music, providing satisfaction through a sense of continuity and organisation in the composition.

It is important to emphasise that the testing user feedback pertaining to musicality, novelty, and coherence serves as an evaluation metric for the algorithm and differs from the user rating utilised as a reward in the Q-Learning algorithm during the training process. Despite the current setup, exploring a reward function incorporating musicality, coherence, and novelty remains a potential avenue for future experimentation.

IV. RESULTS AND EVALUATION

The evaluation encompasses both quantitative metrics and qualitative assessments, with a particular focus on user feedback and the user experience. It is important to note that the testing phase involved a total of 13 user testers, including 3 individuals with expertise in music theory and experience in hobbyist music composition.

*A. Results*

We evaluate the HITL RL music generation algorithm from various angles, including model training evaluation, user interaction and feedback with metric-based assessment.

*1) Model Training Evaluation*

*a) Training Episode Quality*

To assess the quality of musical compositions generated during training episodes, we analysed the trend of user ratings across each episode. Fig. 2 illustrates the improvement in training episode quality over time. Initially, the generated music received low ratings, reflecting the agent's lack of knowledge. However, as training progressed, user ratings steadily increased, indicating improved composition quality. This demonstrates the algorithm's ability to learn and generate better musical content with continuous training.

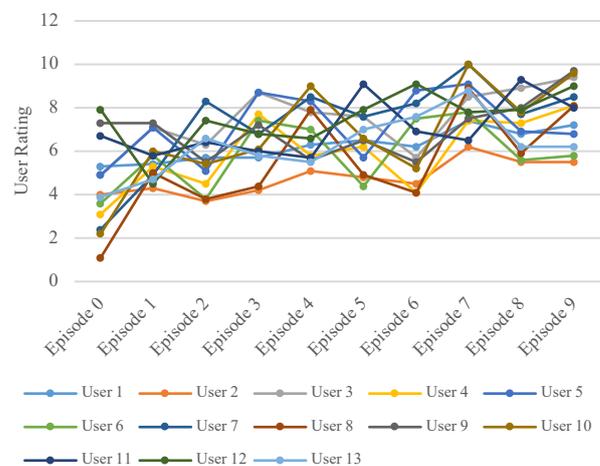

Fig. 2. Training episode quality over time.

*b) Exploration vs. Exploitation*

Examining the logs for exploration strategies, we observed a balanced approach employed by the RL agent when epsilon value, ε= 0.5. Fig. 3 displays the exploration-exploitation balance throughout the training process, affirming a reasonable distribution between exploration and exploitation. The agent effectively explored new musical possibilities during the early stages, gradually shifting towards exploitation as it learned. This balance ensures a diverse range of compositions while leveraging learned knowledge for enhanced quality.

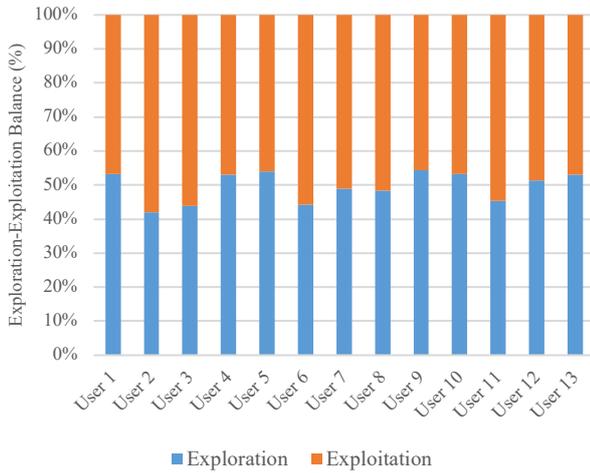

Fig. 3.  Exploration vs. Exploitation

*2) User Interaction and Feedback*

User involvement in the HITL process was crucial for evaluating usability and effectiveness. Users actively engaged with the GUI to generate music and utilised a form for the purpose of furnishing feedback. Fig. 4 illustrates user profiles, depicting their musical expertise by showcasing their familiarity with music theory and experience in music composition.

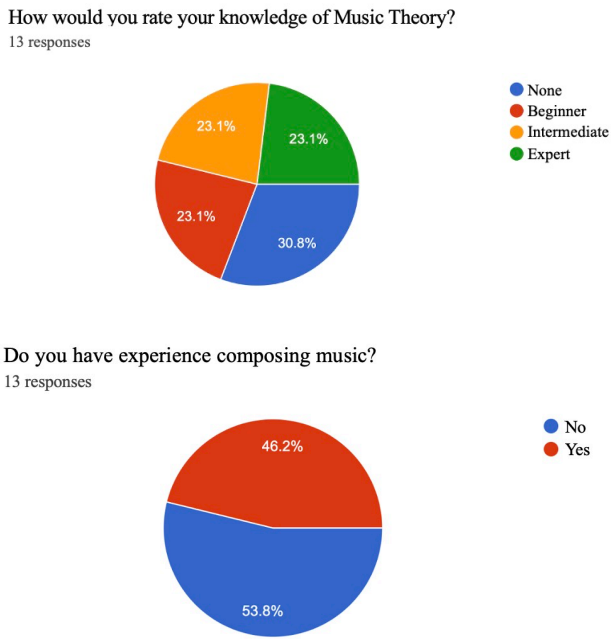

Fig. 4.  User Profiles

*3) Metric-Based Assessment*

User feedback was systematically collected for each generated composition, with testing users providing ratings on a scale of 1-5 for the evaluation metrics outlined in Section III-D: musicality, novelty, and coherence. Fig. 5 illustrates the distribution of user ratings across these three dimensions. The majority of compositions received favourable ratings for musicality and novelty, while coherence ratings showed a wider distribution. This analysis provides insights into areas for potential improvement, with a focus on enhancing novelty in future iterations of the algorithm.

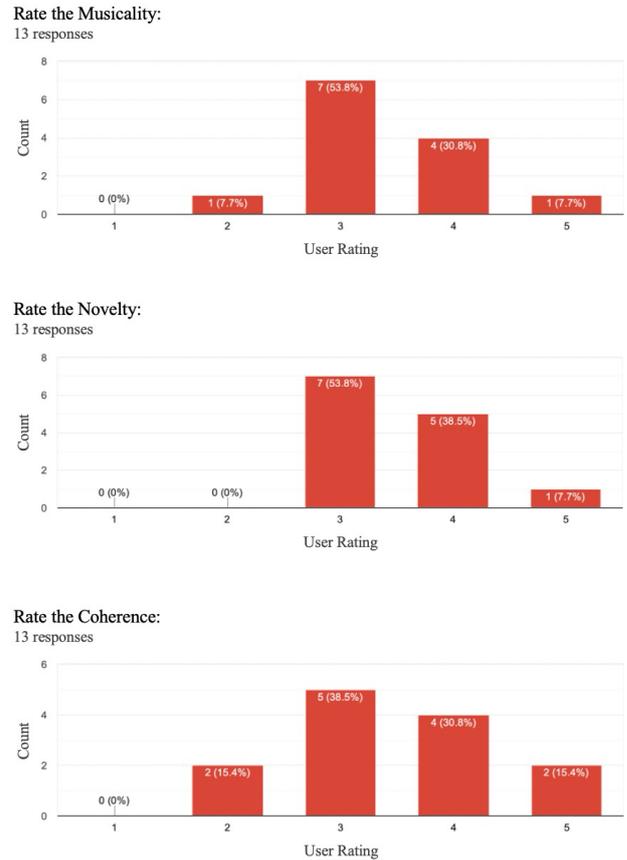

Fig. 5.  User Ratings for Musicality, Novelty, and Coherence

## B. User Experience

The user experience was generally positive. The ability to input preferences such as base note, tempo, volume, and track array length provided control and personalisation.

User feedback was gathered through feedback forms. Participants in the testing process were requested to self-assess their expertise in music theory, selecting from the categories: 'None,' 'Beginner,' 'Intermediate,' and 'Expert.' Additionally, participants were asked about their previous experience in composing music. From the responses to these two inquiries, it was identified that 3 out of the 13 users possessed expertise in music theory and had experience in hobbyist music composition. To account for this distinction in the feedback, we have categorised the user experience feedback into two groups: general users (those lacking music theory expertise or composing experience) and expert users feedback.

*1) General User Feedback*

The feedback from general users indicated a positive user experience. Users praised the melodies being generated, especially when they were informed that no music data was used to train the model. They appreciated the ability to customise preferences like base note, tempo, volume, and track array length, which provided them with control and personalization options.

General users expressed satisfaction with their overall experience, indicating that the system effectively facilitated music composition even for those without expertise in music theory or prior composing experience.

In summary, general users' feedback highlighted the positive aspects of the user experience, emphasising the user-friendliness and the customization options available. This feedback underscores the system's accessibility and effectiveness for a broad user base.

*2) Expert User Feedback*

The two experts in music theory found the model particularly valuable for experimenting with musical ideas and generating raw compositions quickly. However, they emphasised the need for further fine-tuning, especially in the nuances of chord progressions and percussion layering. The expert users' suggestions revolved around several key points:

Enhanced Chord Progressions: Users, including the experts, expressed a desire for more sophisticated and musically rich chord progressions. They noted that the model could benefit from better understanding of harmonic theory and chord resolution.

Dynamic Expression: Users highlighted the importance of adding dynamic expression to the compositions. They suggested incorporating variations in velocity, articulation, and phrasing to make the music more expressive.

Genre Specificity: Some users suggested the incorporation of genre or mood specific modes and constraints. This would enable the model to generate music tailored to specific musical genres or moods.

## V. DISCUSSION

In this section, we delve into a comprehensive discussion of the outcomes, implications, and insights gained from this project on music generation using HITL RL.

### A. Technical Performance

*1) Learning Algorithm and Problem Formulation*

The algorithm discussed in this paper, facilitated by the integration of Q-Learning and Markov Decision Processes (MDP) guided by Bellman's equation, has proven to be a computationally efficient approach. A notable aspect is the user ratings incorporated within each iteration, which, rather than straining computational resources or necessitating advanced algorithms like parallel computing, serve as a manageable bottleneck. While our use of Q-Learning and MDP has yielded the outcomes detailed in Section IV, it would be valuable to consider the implications of alternative reinforcement learning algorithms when combined with HITL input. Such an exploration could shed light on potential enhancements in the quality of the generated music.

*2) The Scale of Algorithmic Music Generation*

The achievement of Q-Learning convergence toward optimal Q-values is contingent on specific conditions. Convergence implies that Q-values stabilise, ceasing significant fluctuations. However, in complex problems, the choice of hyperparameters to reach the optimal policy remains an open question. To determine convergence, we monitor the evolution of Q-values over time and consider the algorithm converged when fluctuations become negligible. It is essential to note that convergence to the optimal policy means obtaining a sequence of actions maximising the agent's reward to reach the goal. For the general problem of music generation, the colossal size of the state space poses an exceptional challenge. For a melody consisting of eight notes within a major scale comprising seven distinct notes, we encountered staggering numbers of potential permutations:

For an 8-note track, $P = 7^8 = 5,764,801$ permutations.

With the introduction of four rhythm variations (out of 8), $P = 7^8 \times 4 = 23,059,204$ permutations.

Further complexity arose when we incorporated two possible percussion pitches, repeated at 4-note intervals and adhering to a consistent rhythm, resulting in an astronomical number of permutations: $P = 7^8 \times 4 \times 2^{4 \times 4} = 1.511208e^{12}$.

These statistics underscore the magnitude of the challenge we confronted when navigating the vast and intricate space of musical possibilities. As outlined in Section V-C, exploring alternative action and state spaces may hold the key to expediting Q-value convergence.

*3) Exploration Strategies*

Our choice of exploration strategy, the Epsilon-Greedy approach, influenced the model's ability to explore new musical ideas. We found that adjusting the epsilon parameter had a significant impact on the model's ability to generate diverse compositions. Fine-tuning this parameter based on user preferences and the stage of training is an area for potential improvement. In addition, evaluating the impact of using other exploration strategies is another area for potential study.

### B. User Experience and Feedback

The incorporation of human feedback through the HITL approach greatly enhanced the user experience. Musicians and composers appreciated the ability to shape the music generation process according to their preferences. This user engagement is a testament to the value of user-guided algorithms in creative domains.

### C. Future Directions

*1) Model Complexity*

Expanding the complexity of the HITL RL model, possibly by incorporating more advanced RL algorithms or neural architectures, could lead to further improvements in music generation. Additionally, exploring generative models like GANs (Generative Adversarial Networks) could open new avenues for creativity.

*2) Collaboration and Sharing*

Encouraging collaboration among musicians and composers by allowing them to share their user-tailored models could foster innovation and the creation of new musical genres.

*3) Alternative State Space*

One avenue for enhancing the likelihood of convergence involves the exploration of alternative state space representations. As previously discussed in Section V-A, the potential permutations of tracks are in the millions. Presently, the algorithm uses the *track array* as the state. However, it may prove advantageous to experiment with a novel representation of the *track array*, one that is relative to the base note or employs another form of representation. Such an approach could potentially reduce the size of the state space, thereby promoting faster convergence of Q-values.

*4) Alternative Reward Function*

Currently, the reward of generated music relies on the human input in the form of a user rating on a scale of 1 to 10. Looking ahead, a promising avenue for future exploration involves experimenting with a more nuanced and comprehensive approach to user feedback. Specifically, the introduction of a compound reward function, intricately tied to user feedback rating on novelty, musicality, and coherence, as outlined in Section III-D. By incorporating expert assessments into the reward function, the algorithm could potentially evolve to produce music that aligns more closely with expert standards and nuanced artistic criteria.

## VI. CONCLUSION

In conclusion, this paper introduces a novel approach to algorithmic music generation by combining Human-In-The-Loop Reinforcement Learning (HITL RL) with principles derived from music theory. It represents a significant step toward user-guided algorithmic music composition. We have demonstrated the feasibility of combining reinforcement learning with user interaction to create original and customizable musical compositions. The technical performance, user experience, and potential future avenues are all integral aspects that will shape the future of this field. As we continue to refine our approach and address challenges, we look forward to contributing to the evolving landscape of algorithmic music generation.

This study represents a significant stride in user-guided algorithmic music composition. Despite persisting challenges, such as achieving nuanced chord progressions and dynamic expression, the model facilitates experimentation and creativity in music composition. As we refine our approach based on user feedback and expert input, we anticipate further advancements in algorithmic music generation capabilities. The HITL RL music generation algorithm holds immense potential for musicians, offering a collaborative platform for music creation and exploration without relying on pre-existing musical data. This ongoing exploration aligns with the evolving landscape of algorithmic music generation, contributing to the intersection of technology and artistic expression.